\documentclass[pre,showpacs,twocolumn,twoside,aps,a4paper,groupedaddress,floatfix]{revtex4}
\usepackage{t1enc}
\usepackage{amsmath,amssymb,mathtools}
\usepackage{hhline,bbm,psfrag}
\usepackage{subfigure}
\usepackage{graphicx}

\newcommand{\pd}{\partial}
\newcommand{\dt}{{\Delta t}}

\newcommand{\bex}{\mathbf{e_x}}
\newcommand{\bey}{\mathbf{e_y}}
\newcommand{\bez}{\mathbf{e_z}}
\newcommand{\lap}{\nabla^2}
\newcommand{\lapd}{\nabla^4}

\newcommand{\p}{^\prime}
\newcommand{\Temp}{\Theta}
\newcommand{\Tempb}{\overline{\Temp}}

\newcommand{\Conc}{C}
\newcommand{\Concb}{\overline{\Conc}}

\newcommand{\Strm}{\Psi}

\newcommand{\Psib}{\overline{\Strm}}

\newcommand{\bU}{\mathbf{U}}
\newcommand{\bUb}{\mathbf{\overline{U}}}
\newcommand{\TW}{{\rm TW}}
\newcommand{\SW}{{\rm SW}}
\newcommand{\bSW}{{\overline{\SW}}}

\newcommand{\bu}{\mathbf{u}}

\newcommand{\mL}{\mathcal{L}}
\newcommand{\mJ}{\mathcal{J}}
\newcommand{\mN}{\mathcal{N}}
\newcommand{\domain}{\lambda}
\newcommand{\Tper}{T}
\newcommand{\omegacond}{\omega_{\rm cond}}
\newcommand{\sigmacond}{\sigma_{\rm cond}}

\newcommand{\omegatwmean}{\omega^{\rm TW}_{\rm mean}}
\newcommand{\sigmatwmean}{\sigma^{\rm TW}_{\rm mean}}
\newcommand{\omegaswmean}{\omega^{\rm SW}_{\rm mean}}
\newcommand{\sigmaswmean}{\sigma^{\rm SW}_{\rm mean}}
\newcommand{\omegatwobs}{\omega^{\rm TW}_{\rm exact}}
\newcommand{\omegaswobs}{\omega^{\rm SW}_{\rm exact}}
\newcommand{\sigmamean}{\sigma_\bUb}

\newcommand{\omegamean}{\omega_\bUb}
\newcommand{\omegaobs}{\omega}
\newcommand{\Ref}{\Pi^{\rm Ref}}
\newcommand{\Trans}{\Pi^{\rm Trans}}
\newcommand{\Time}{\Pi^{\rm Time}}
\newcommand\Bouss{\Pi^{\rm Bouss}}
\begin{document}
\title{Prediction of frequencies in thermosolutal convection 
from mean flows}
\author{Sam E. Turton}
\affiliation{DAMTP, Centre for Mathematical Sciences, Wilberforce Road, Cambridge CB3 0WA, United Kingdom}
\author{Laurette S. Tuckerman}
\affiliation{PMMH (UMR 7636 CNRS - ESPCI - UPMC Paris 6 - UPD Paris 7 - ParisTech - PSL)\\
  10 rue Vauquelin, 75005 Paris, France}
\author{Dwight Barkley}
\affiliation{Mathematics Institute, University of Warwick, 
Coventry CV4 7AL, United Kingdom}

\date{\today}

\begin{abstract}
Motivated by studies of the cylinder wake, in which the vortex-shedding 
frequency can be obtained from the mean flow, 
we study thermosolutal convection driven by opposing thermal and 
solutal gradients. In the archetypal two-dimensional geometry 
with horizontally periodic and vertical no-slip boundary conditions, 
branches of traveling waves and standing waves are created 
simultaneously by a Hopf bifurcation. 
Consistent with similar analyses performed on the cylinder wake, 
we find that the traveling waves of thermosolutal convection
have the RZIF property, meaning that 
linearization about the mean fields of the 
traveling waves yields an eigenvalue whose real part is almost zero 
and whose imaginary part corresponds very closely to the 
nonlinear frequency.
In marked contrast, linearization about the mean field of the standing 
waves yields neither zero growth nor the nonlinear frequency.
It is shown that this difference can be attributed to the fact that 
the temporal power spectrum for the traveling waves is peaked, 
while that of the standing waves is broad. We give a general demonstration 
that the frequency of any quasi-monochromatic oscillation 
can be predicted from its temporal mean.

\end{abstract}

\pacs{
47.20.Ky, 
47.20.Bp, 
}

\maketitle

\section{Introduction}

One of the most important characterizations of an oscillating system is its
frequency. 
In fluid dynamics, perhaps the best known example of a oscillating system is
the von K\'{a}rm\'{a}n vortex street generated in the wake of a circular
cylinder. Because this oscillating flow arises from a supercritical Hopf
bifurcation, the frequencies observed in experiments~\cite{Williamson} and in
direct numerical simulations~\cite{Dusek} agree at onset with the Hopf
frequency obtained from a linear stability analysis of the steady
flow~\cite{Jackson}. However, as the oscillations grow in amplitude away from
the bifurcation, their frequencies differ substantially from those obtained 
from linear stability analysis, and as a result, even quite close to onset,
standard stability analysis of steady base flows dramatically fails to predict 
observed oscillation frequencies.



This has led to a growing interest in obtaining frequencies of 
oscillating systems, the cylinder wake in particular, 
by analysing their temporally-averaged profiles,
rather than steady base flows. 
Hammond and Redekopp \cite{Redekopp}, 
Pier \cite{Pier}, Barkley \cite{Barkley}, and Mittal \cite{Mittal} 
have shown that a linear stability analysis about the mean velocity profile 
yields an eigenvalue whose imaginary part 
corresponds very closely to the nonlinear frequency. 
In addition, the real part of the eigenvalue is virtually zero, 
which Barkley \cite{Barkley} 
interpreted as the marginal stability of the mean flow.
This property, which we shall call the real-zero imaginary-frequency, or RZIF
property, will be of central interest in the following.



The importance of the mean field and of its marginal stability 
was the subject of classic articles by Malkus \cite{Malkus_56}
and by Stuart \cite{Stuart_58}. 
Wesfreid and co-workers have measured the the mean flow in 
configurations such as 
the wake of a triangular obstacle \cite{Zielinska} 
and a confined jet \cite{Maurel}.
Noack \cite{Noack} have formulated a hierarchy of low-dimensional models 
for the cylinder wake which incorporate 
both the mean flow and the nonlinear limit cycle. 


RZIF was further studied 
by Sipp and Lebedev \cite{Sipp} by means of a weakly nonlinear expansion 
about the Hopf bifurcation point, in which the successive  
two-dimensional solvability conditions and eigenproblems generated 
at each order were solved numerically.
%
They demonstrated the non-universality of RZIF
by carrying out the procedure for both the cylinder wake and the oscillatory 
flow over a square cavity and showing that, close to onset, 
the cylinder-wake mean flow had the RZIF property while the cavity 
mean flow did not. 
Mantic-Lugo {\em et al}.~\cite{Gallaire} showed that 
the assumption of RZIF could be used to calculate an accurate approximation to 
the mean flow without the need to compute the nonlinear limit cycle. 
More specifically, they calculated a flow 
such that linearization about it 
led to an eigenvalue with zero real part, and 
showed that this flow was extremely close to the actual 
mean flow of the limit cycle. 

Until now, RZIF has been investigated only for open flows 
and almost exclusively the cylinder wake.
Here, we carry out a similar investigation of the traveling and standing waves 
which emerge simultaneously from a Hopf bifurcation in thermosolutal 
convection. 
We find that the traveling waves are an ideal case of RZIF;
the standing waves, however, do not display this phenomenon at all. 

Finally, and most significantly, 
we then show that RZIF is closely connected to the temporal spectrum of the
nonlinear oscillations. 
If the temporal dependence is monochromatic, i.e. if the oscillations 
are trigonometric, then RZIF is exactly satisfied. 
If the spectrum decays rapidly away from the main frequency, 
as is the case for the thermosolutal traveling waves but not 
the standing waves, then RZIF is approximately satisfied. 


\section{Thermosolutal convection}

Thermosolutal convection is driven by independently 
imposed gradients in the temperature and concentration of the fluid. 
More specifically, temperatures and concentrations are set 
to values which differ by $\Delta \Temp$ and $\Delta C$
at two plates separated by a vertical distance $h$.
Under the Boussinesq approximation, the density $\rho$ in the 
buoyancy term is assumed to depend linearly on temperature and concentration,
with coefficients $\rho_T$ and $\rho_C$; the kinematic viscosity $\nu$ 
and thermal and solute diffusivities $\kappa_T$ and $\kappa_C$ 
are assumed to be constant.
We nondimensionalize lengths, temperature, concentration and time by
$h$, $\Delta \Temp$, $\Delta C$ and the thermal viscous time $h^2/\kappa_T$. 
One solution to the equations of motion is the conductive state,
in which the fluid is motionless and the thermal and solutal profiles 
vary linearly with height; we 
denote by $\Temp$ and $\Conc$ the deviations of temperature and 
concentration from the conductive state. 
Restricting ourselves to the two-dimensional case, 
a streamfunction $\Psi$ is used to represent the velocity as 
$\nabla\Psi\times\bey = -\pd_z\Psi \bex + \pd_x \Psi\bez$. 
The governing equations are then:
\begin{subequations}
\begin{align}
\pd_t \Temp - \mJ[\Psi,\Temp] &= \lap \Temp + \pd_x \Psi\\
\pd_t C - \mJ[\Psi,C] &= L\lap C + \pd_x \Psi\\
\label{eq:goveqvort}
\pd_t \lap \Psi - \mJ[\Psi, \lap\Psi]  &=P\left(\lapd \Psi + \pd_x\left(R_T \Temp+R_S C\right)\right)
\end{align}
\label{eq:goveq}
\end{subequations}
where the Poisson bracket is
\begin{align}
\mJ[f,g] &\equiv \bey\cdot\nabla f\times\nabla g = \pd_zf \pd_xg -\pd_xf \pd_zg 
\label{eq:Poisson}\end{align}
and where the nondimensional parameters in \eqref{eq:goveq} 
are the Prandtl and Lewis numbers
\begin{align}
P =\frac{\nu}{\kappa_T} \qquad L = \frac{\kappa_C}{\kappa_T}
\end{align}
and the thermal and solutal Rayleigh numbers
\begin{align}
R_T = \frac{g\rho_T \Delta \Temp h^3}{\nu\kappa_T} \qquad
R_S = \frac{g\rho_C \Delta C h^3}{\nu\kappa_T} 
\label{eq:rtrs}\end{align}
(Note that according to the conventions used here, 
$\kappa_T$ is present in both of the denominators 
in \eqref{eq:rtrs}.)
Our study concerns the waves that result when $R_T$ and $R_S$ 
are of opposite signs. 

By defining
\begin{align}
\bU \equiv(\Temp, C, \lap\Psi)^T
\end{align}
we can rewrite \eqref{eq:goveq} in the compact notation
\begin{align}
\pd_t\bU = \mL \bU + \mN(\bU,\bU)
\label{eq:compact}\end{align}
where $\mN$ corresponds to the quadratic terms of the Poisson bracket 
$\mJ$ on the left-hand-side of \eqref{eq:goveq} and 
$\mL$ denotes the terms on the right-hand-sides of \eqref{eq:goveq}, 
which are linear.

We use the simplest possible boundary conditions, 
namely periodic boundary conditions in the horizontal direction
\begin{subequations}
\begin{align}
\bU(x+\domain,z)=\bU(x,z)
\label{eq:xperiodic}\end{align}
and fixed temperature and concentration at free-slip boundaries 
with no horizontal flux
\begin{align}
\bU(x,z=0) = \bU(x,z=1) 
\label{eq:freeslip}\end{align}
\label{eq:bcs}
\end{subequations}
\noindent where conditions \eqref{eq:xperiodic} and \eqref{eq:freeslip} are 
imposed on $\Psi$ as well as $\lap\Psi$, $T$ and $C$.
The assumption of free-slip boundaries 
simplifies the problem while reproducing qualitatively 
the essential features of thermosolutal convection.

The pure thermal problem \eqref{eq:compact}-\eqref{eq:bcs} with $R_S=0$
undergoes a bifurcation to stationary convection, 
whose threshold in the free-slip case 
was found by Rayleigh~\cite{Rayleigh} to be minimized by 
the classic critical wavenumber and Rayleigh number 
\begin{align}
k=\frac{\pi}{\sqrt{2}} \qquad
R_0=\frac{(k^2+\pi^2)^3}{k^2}\equiv\frac{q^6}{k^2}
\label{eq:crit}\end{align}
Based on \eqref{eq:crit}, 
we use as the periodicity length $\domain\equiv 2\pi/k = 2\sqrt{2}$ and
we define the reduced Rayleigh number and separation parameter:
\begin{align}
r\equiv \frac{R_T}{R_0} \qquad S\equiv \frac{R_S}{R_T} 
= \frac{R_S}{r R_0}
\end{align}
so that 
\begin{align}
R_T \Temp+R_S C=r R_0(\Temp+SC)
\label{eq:rtsc}\end{align}
in \eqref{eq:goveqvort}. The case of interest to us is $S<0$.

In what follows, we set $P=10$, $L=0.1$ and $S=-0.5$. 
Results will concern either the interval $2\leq r \leq 3$
or else the specific value $r=2.5$.

\section{Bifurcations and Symmetry}
\label{sec:bifs}

System \eqref{eq:compact}-\eqref{eq:bcs} undergoes 
a number of bifurcations. 
For $S$ in the range $-1 \lesssim S \lesssim -L^2$, 
a Hopf bifurcation occurs at $r=r_H$ with Hopf frequency $\omega_H$.
These parameters have particularly 
simple expressions \cite{Tuck_PhysD} for $P\gg 1$, $L\ll 1$:
\begin{align}
r_H=\frac{1}{1+S} \qquad
{\omega_H}^2=-q^4\frac{S}{1+S}
\end{align}
For our case, with $P=10$, $L=0.1$, and $S=-0.5$, we have 
\begin{align}
r_H=2.05 \qquad \omega_H=13.5
\end{align}
close to the $P\gg 1$, $L\ll 1$ values of 2 and $q^2\approx 14$.

The system \eqref{eq:compact}-\eqref{eq:bcs} has $O(2)$ symmetry in $x$, 
i.e. it is invariant under the translation and reflection operators:
\begin{align}
\Trans_{x_0} (\Temp,C,\Psi)(x,z,t) \equiv (\Temp,C,\Psi)(x+x_0,z,t) 
\label{eq:Trans}\\
\Ref (\Temp,C,\Psi)(x,z,t) \equiv (\Temp,C,-\Psi)(-x,z,t) 
\label{eq:Ref}\end{align}
It also has an additional $Z_2$ symmetry, namely the Boussinesq symmetry 
which combines reflection in $z$ with reversal of the sign of 
temperature and concentration deviations:
\begin{align}
\Bouss (\Temp,C,\Psi)(x,z,t) \equiv (-\Temp,-C,-\Psi)(x,1-z,t) 
\label{eq:Bouss}\end{align}
Finally, \eqref{eq:compact}-\eqref{eq:bcs} is invariant under time translation:
\begin{align}
\Time_{\dt} (\Temp,C,\Psi)(x,z,t) \equiv (\Temp,C,\Psi)(x,z,t+\dt) 
\end{align}
Knobloch \cite{Knobloch_86} has shown that a Hopf bifurcation leading to 
the breaking of 
translation symmetry within $O(2)$ leads to branches of traveling waves (TW) 
and standing waves (SW), at most one of which is stable.
For the parameter values we use, both TW and SW bifurcate 
supercritically with increasing $r$ and it is the TW branch which is stable.

\begin{figure}
\includegraphics[width=0.8\columnwidth]{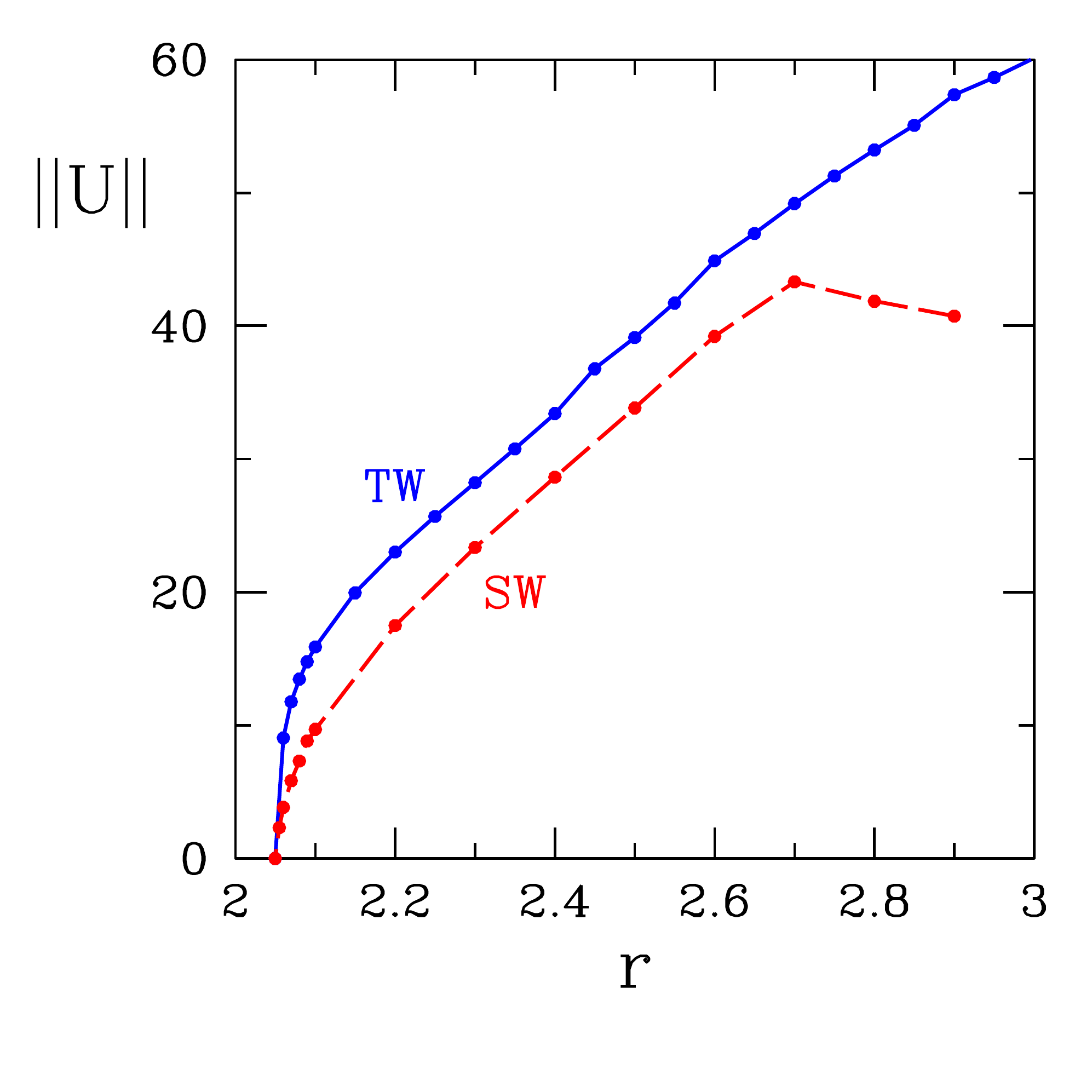}
\caption{(Color online) Branches of traveling and standing wave
  states.  The higher-amplitude TW branch (blue, solid) is stable near
  onset, while the SW branch (red, dashed) is unstable.  The
  bifurcation to TW is degenerate, so that $||U||\sim (r-r_H)^{1/4}$.
  The SW branch may undergo a secondary bifurcation at $r\approx
  2.7$.}
\label{fig:bifdiag}
\end{figure}
A bifurcation diagram showing the amplitudes $||U||$ where
\begin{align}
||U||^2 \equiv \int_0^\lambda dx \: dz \: 
&\left(|\Temp(x,z,t)|^2 \right.\nonumber\\ 
+ &\left. |\Conc(x,z,t)|^2 + |\Psi(x,z,t)|^2\right]
\end{align}
of the traveling waves and the standing waves is shown in figure 
\ref{fig:bifdiag}.
For TW, $||U||$ is independent of time and for SW, 
$||U||$ is evaluated at the moment at which the integral 
of the temperature is maximal.
The abrupt onset of the TW branch reflects
the fact that the Hopf bifurcation to traveling waves 
in thermosolutal convection with free-slip boundary conditions 
is degenerate \cite{Knobloch_85,Knobloch_89}, so that $||U||\sim (r-r_H)^{1/4}$ 
rather than the usual square-root dependence. The non-monotonic 
behavior of the amplitude of the SW branch may indicate 
a secondary transition, which will not concern 
us in this investigation.
The traveling waves are invariant under the combined Boussinesq-shift symmetry:
\begin{align}
\Bouss\Trans_{\domain/2} &(\Temp,C,\Psi)(x,z,t)\nonumber\\
= &(-\Temp,-C,-\Psi)(x+\domain/2,1-z,t)
\label{eq:BoussTrans}\end{align}
The TW are also invariant under the family of spatio-temporal symmetries
\begin{align}
\Time_\dt\Trans_{-v\dt}&(\Temp,C,\Psi)(x,z,t)\nonumber\\
= &(\Temp,C,\Psi)(x-v\dt,z,t+\dt)
\end{align}
where $v\equiv\domain/T = \omega/k$ is the velocity.
Thus, the TW have $SO(2)\times Z_2$ symmetry.
The standing waves are invariant at all times under 
$\Ref$ and $\Bouss$, 
i.e. the SW have $Z_2 \times Z_2$ symmetry. 

\section{Linearization and mean fields}

The usual linear stability analysis problem about the conductive state 
is written as 
\begin{align}
\pd_t \bu = \mL \bu
\label{eq:classiclinstab}\end{align}
where the infinitesimal perturbation is
\begin{align}
\bu \equiv (\tau,c,\lap\psi)^T.
\end{align}
We recall that $\mL$ encompasses the terms on the right-hand-side of 
\eqref{eq:goveq}. 

A Hopf bifurcation from the conductive state takes place when 
the real part of an eigenvalue $\sigmacond\pm i \omegacond$ of $\mL$ crosses 
zero. 
Because of the periodic and no-slip boundary conditions \eqref{eq:bcs}, 
eigenvectors of $\mL$ are of the simple spatial form
\begin{align}
\left(\begin{array}{c}\Temp_0 \sin(kx)\sin(\pi z)\\ C_0\sin(kx) \sin(\pi z)\\ 
\Psi_0 \cos(kx)\sin(\pi z) \end{array}\right),
\left(\begin{array}{c}\Temp_0 \cos(kx)\sin(\pi z)\\ C_0\cos(kx) \sin(\pi z)\\ 
\Psi_0 \sin(kx)\sin(\pi z) \end{array}\right)
\label{eq:evecs0}\end{align}
and linear combinations of these vectors, 
where $\Temp_0$, $C_0$, $\Psi_0$ are complex scalars. 
Complex eigenmodes of $\mL$ 
are associated with a four-dimensional eigenspace. Some combinations 
of eigenmodes lead naturally to standing waves, others to traveling waves 
\cite{Boronska}.

Our study concerns the temporal means of the traveling and 
standing waves. We define
\begin{align}
\bUb(x,z)\equiv\frac{1}{\Tper}\int_{t=0}^{\Tper} \bU(x,z,t) dt
\label{eq:Tmean}\end{align}
where $\Tper$ is the temporal period. 
Because the temperature and concentration are important 
components of $\bUb$, we refer to mean fields rather than to mean flows, 
contrary to the purely hydrodynamic literature.
The mean field $\bUb$ is not a solution of the governing equations. 
Averaging the governing equations \eqref{eq:compact}, we obtain
the equation obeyed by $\bUb$, which is
\begin{align}
0 = \mL \bUb + \overline{\mN(\bU,\bU)}= \mL \bUb + \mN(\bUb,\bUb) + F 
\label{eq:RANS}\end{align}
where 
\begin{align}
F \equiv \overline{\mN(\bU-\bUb,\bU-\bUb)}
\end{align}
is analogous to the usual Reynolds stress force for the Navier-Stokes 
equations \cite{Barkley,Gallaire}
and can be interpreted as the force that would have to be exerted 
on the system in order for the mean field to be a steady solution.

We now linearize the governing equations \eqref{eq:compact} 
about $\bUb$:
\begin{align}
\pd_t\bu = \mL \bu + \mN(\bUb,\bu) + \mN(\bu,\bUb) \equiv \mL_\bUb\bu
\end{align}
This procedure is not a conventional linear stability analysis 
since $\bUb$ is not a steady solution unless $F=0$. 
For clarity, when 
$\bUb$ is the mean field of nonlinear SW or TW states, 
the operator $\mL_\bUb$ will be denoted by $\mL_\TW$ or $\mL_\SW$ and its  
leading eigenvalues will be denoted by 
$\sigmatwmean \pm i \omegatwmean$ or $\sigmaswmean \pm i \omegaswmean$. 
These eigenmodes are the main focus of the following sections. 

\section{Traveling waves}
\label{sec:TW}

\begin{figure*}
\centerline{
\includegraphics[width=2\columnwidth,clip]{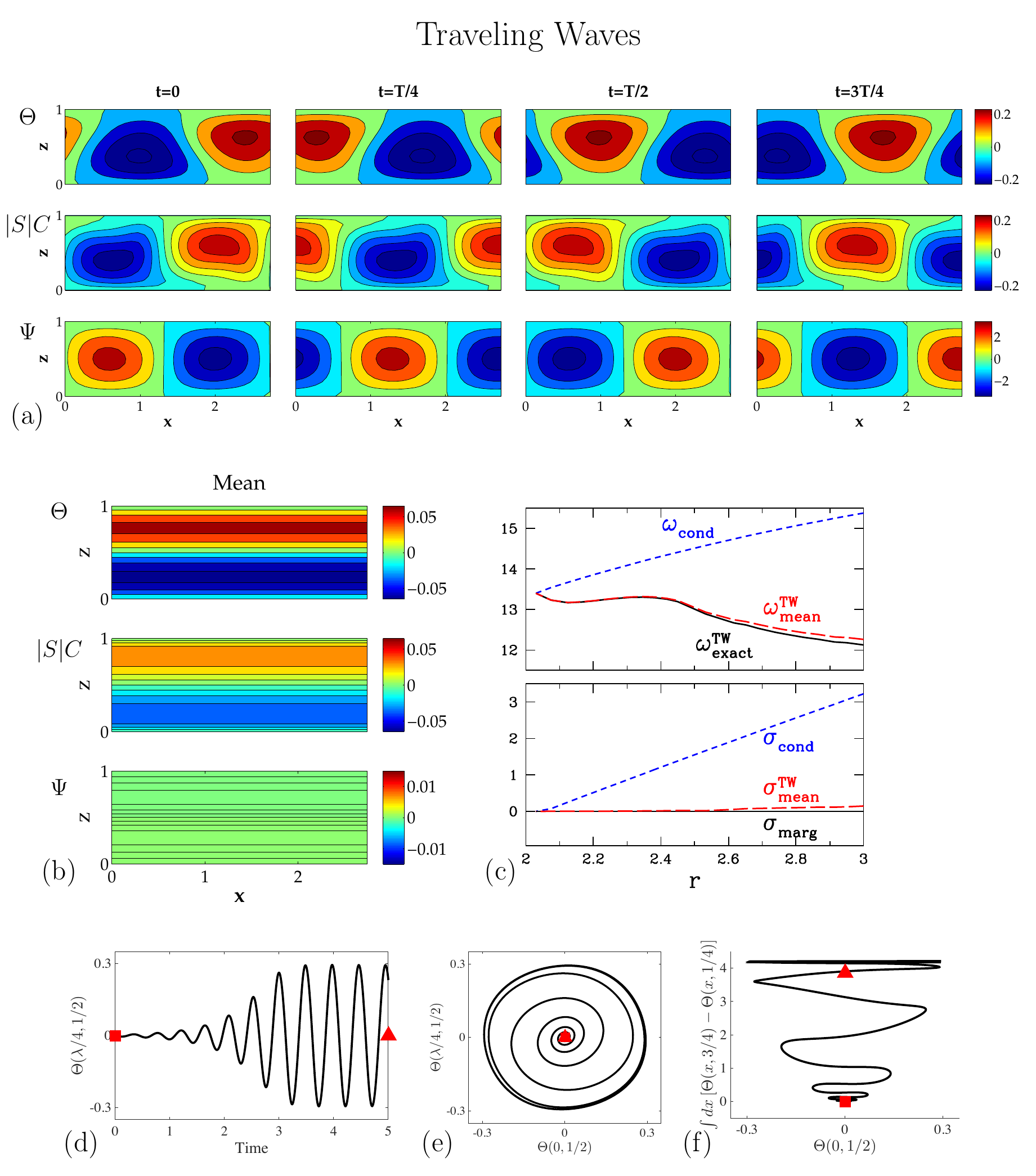}}
\caption{(Color online) (a) Instantaneous temperature, concentration and streamfunction 
for a traveling wave at $P=10$, $L=0.1$, $S=-0.5$ and $r=2.5$.
$\Temp$ is out of phase with $\Conc$ and $\Psi$.
The Boussinesq-shift symmetry \eqref{eq:BoussTrans} is especially clear for $\Conc$.
Endpoints of ranges of $\Temp$, $\Conc$, $\Psi$ are $\pm 0.32$, $\pm 0.46$, 
and $\pm 3.4$, respectively.
(b) Temporally averaged fields $\Tempb$, $\Concb$, $\Psib$.
for this traveling wave. The functional form is 
approximately $\sin(2\pi z)$ and the amplitude of $\Psib$ 
is much smaller than that of $\Tempb$, $\Concb$.
Endpoints of ranges of $\Temp$, $\Conc$, $\Psi$ are $\pm 0.065$, 
$\pm 0.077$ and $\pm 1.7 \times 10^{-4}$, respectively.
(c) Frequencies (above) and growth rates (below) as a 
function of Rayleigh number for traveling waves. 
The observed frequency ($\omegatwobs$, solid black) is closely tracked by the 
frequency of the traveling wave mean field ($\omegatwmean$, long-dashed red) 
and not at all by 
the frequency of the conductive state ($\omegacond$, short-dashed blue).
The growth rate $\sigmatwmean$ is near zero, indicating that 
$\bUb_\TW$ is marginally stable.
The horizontal line $\sigma_{\rm marg}=0$ is used to indicate 
marginal stability.
(d) Timeseries of $\Temp(x=\lambda/4,z=1/2)$ for $r=2.5$
from low-amplitude initial condition. 
Projection of conductive state is indicated by a square and of 
the mean field by a triangle. 
(e) Phase portrait showing projection of time-dependent evolution 
onto the temperature at $(x=0,z=1/2)$ and $(x=\lambda/4,z=1/2)$.
(f) Projection onto $\Temp(x=0,z=1/2)$ 
and $P_M$ of \eqref{eq:meanproxy}, 
which is a proxy for projection onto the mean field. }
\label{fig:TWfields}
\end{figure*}

\begin{figure*}
\centerline{
\includegraphics[width=2\columnwidth,clip]{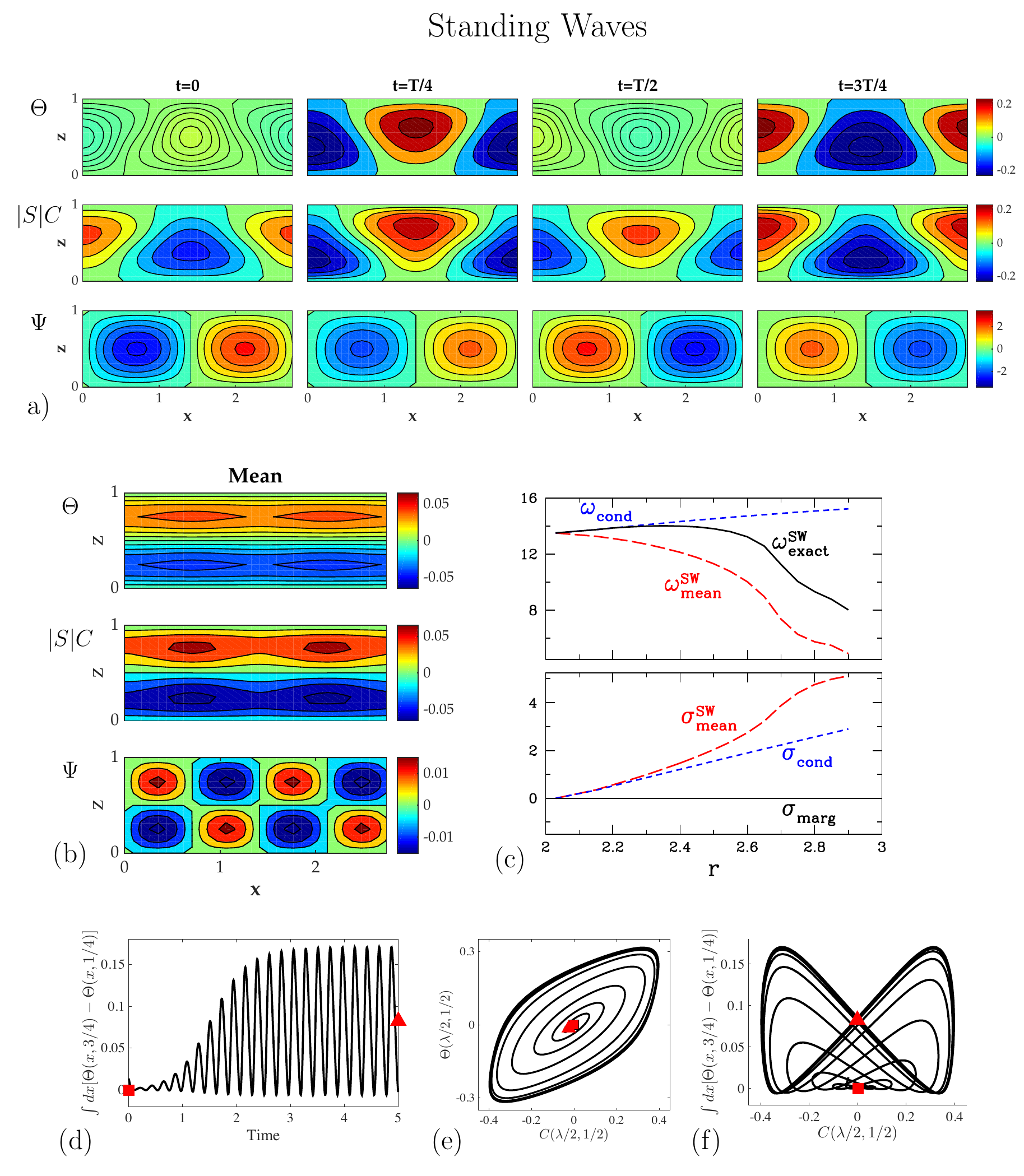}}
\caption{(Color online) (a) Instantaneous temperature, concentration and streamfunction 
for a standing wave at $P=10$, $L=0.1$, $S=-0.5$ and $r=2.5$.
The spatial features, particularly the nodal curves, of $\Temp$ and $\Conc$ 
are very similar, but there is a $T/8$ temporal shift between the two.
These fields have both the Boussinesq-shift symmetry \eqref{eq:BoussTrans} 
and $x$-reflection symmetry \eqref{eq:Ref}.
Endpoints of ranges of $\Temp$, $\Conc$, $\Psi$ are $\pm 0.33$, $\pm 0.46$, 
and $\pm 2.6$, respectively.
The color scale is the same as that of figure \ref{fig:TWfields}(a).
(b) Temporally averaged fields $\Tempb$, $\Concb$, $\Psib$ 
for this standing wave. 
Endpoints of ranges of $\Temp$, $\Conc$, $\Psi$ are $\pm 0.043$, $\pm 0.13$, 
and $\pm 0.016$, respectively.
The color scale is the same as that of figure \ref{fig:TWfields}(b).
(c) Frequencies (above) and growth rates (below) as a 
function of Rayleigh number for standing waves.
The observed frequency ($\omegaswobs$, solid black) is initially 
tangent to that of the conductive state ($\omegacond$, short-dashed blue), 
and then, at $r \approx 2.4$, veers downwards. 
The frequency of the standing wave mean field ($\omegaswmean$,
long-dashed red) deviates significantly from the observed frequency almost 
immediately after onset, although it shows a similar downward trend.
The growth rate $\sigmaswmean$ is never small and so the 
mean field $\bUb_\SW$ is never marginal.
(d) Timeseries of $P_M$ of \eqref{eq:meanproxy} for $r=2.5$
from low-amplitude initial condition. 
Projection of conductive state is indicated by a square and of 
the mean field by a triangle. 
(e) Time-dependent evolution projected onto the temperature 
and the concentration at the same point $(x=\lambda/2,z=1/2)$.
(f) Projection onto $\Conc(x=\lambda/2,z=1/2)$ and $P_M$.}
\label{fig:SWfields}
\end{figure*}

Figure \ref{fig:TWfields}(a) shows snapshots of a traveling wave $\bU_{\rm TW}$.
We show deviations $\Theta$ and $\Conc$ from the conductive temperature 
and concentration profiles and the streamfunction $\Psi$ of the velocity field. 
Equation \eqref{eq:rtsc} shows that $|S|\Conc$ is the appropriate 
quantity to compare with $\Temp$.

Figure \ref{fig:TWfields}(b) shows the mean field $\bUb_{\rm TW}$, 
of this traveling wave. 
For traveling waves, this temporal average 
can be obtained by an instantaneous spatial average:
\begin{align}
\bUb_{\rm TW}(z) &\equiv \frac{1}{\Tper}\int_0^{\Tper}
\:dt\: \bU_{\rm TW}(x-v t,z)\nonumber\\
&= \frac{1}{\domain}\int_0^{\domain}\:dx\: \bU_{\rm TW}(x-v t,z)
\end{align}
and thus is necessarily independent of $x$.
The mean fields are entirely different 
from the cellular instantaneous fields:
the instantaneous fields are dominated by 
periodic dependence in $x$ which vanishes 
upon integration and a much smaller $x$-independent component. 

If $\bU$ were of the same trigonometric form as the eigenvectors
\eqref{eq:evecs0}, its spatial and temporal average 
would necessarily be zero. 
Nonlinear effects are responsible for modifying the 
spatial dependence of the traveling waves and creating the mean field.
The form of the mean fields can be explained 
heuristically by 
the lowest order nonlinearity, namely substituting the spatial 
form \eqref{eq:evecs0} of the eigenvectors into the nonlinear term:
\begin{align}
\mJ[\Psi,\Temp] &= 
\mJ[\Psi_0 \cos(kx) \sin(\pi z),\Temp_0 \sin(kx)\sin(\pi z)] \nonumber\\
&= k\pi\Psi_0\Temp_0 \sin(2\pi z)/2
\label{eq:sin2}\end{align}
and similarly for $\mJ[\Psi,C]$.
Since in \eqref{eq:evecs0}, the streamfunction is an eigenfunction of 
the Laplacian, the lowest order nonlinear contribution to the 
streamfunction equation vanishes:
\begin{align}
\mJ[\Psi,\lap\Psi] = \mJ[\Psi,-q^2\Psi] = 0
\label{eq:psijpsi}\end{align}
Indeed, the mean fields shown in figure \ref{fig:TWfields}(b) have a functional 
form like \eqref{eq:sin2} and the amplitude of the mean streamfunction 
is very small: $||\Psib||/||\Psi|| \sim 5\times 10^{-5}$, compared to 
$||\Tempb||/||\Temp|| \sim ||\Concb||/||\Conc|| \sim 0.2$, 
again motivating our use of the term mean field, rather than mean flow.
%
The observations 
\eqref{eq:sin2} and \eqref{eq:psijpsi} are 
those which lead to the formulation of the three-variable 
Lorenz model for convection \cite{Lorenz} and the five-variable Veronis 
model for thermosolutal convection \cite{Veronis_65}.

Figure \ref{fig:TWfields}(c) constitutes one of our main findings,
namely that linearization about the mean field of the TWs yields an
eigenvalue whose imaginary part matches the frequency of the nonlinear
traveling waves and whose real part is close to zero.
That is, the traveling waves in thermosolutal convection have the 
RZIF property, which is manifested in the figure by the 
fact that the long-dashed red and solid black curves coincide. 
The real and imaginary parts, $\sigmatwmean$ and $\omegatwmean$, 
of the leading eigenvalues of $\mL_\TW$ are shown. 
The frequency $\omegatwmean$ agrees
almost exactly with the actual frequency $\omegatwobs$ of the 
nonlinear traveling waves over the entire range of our study, $2 < r < 3$.
Furthermore, the growth rate $\sigmatwmean$ of this eigenvalue remains
approximately zero up to at least 50\(\%\) above onset. This implies
that the mean fields, viewed as solutions of the governing
equations for thermosolutal convection - with the appropriate
Reynolds stress forcing - are marginally stable.

Figure \ref{fig:TWfields}(c) also shows $\sigmacond$ and $\omegacond$, 
the real and imaginary parts of the leading eigenvalue of $\mL$. 
Since the onset of traveling waves corresponds to a 
supercritical Hopf bifurcation 
of the conductive state, $\sigmacond$ crosses zero and 
the non-zero frequency $\omegacond$ necessarily agrees with the 
nonlinear frequency at onset. 
However, past the bifurcation, the frequencies diverge from one other 
and $\omegacond$ drastically overpredicts the observed value 
$\omegatwobs$. 
This is very much like the situation for the cylinder wake~\cite{Barkley}.

Figure \ref{fig:TWfields}(d)-(f)
illustrates the relationship between the traveling wave $\bU_\TW$, 
its mean field $\bUb_\TW$, and the conductive state. 
The time series, figure \ref{fig:TWfields}(d), shows the 
temperature at a representative point. The field at $t=0$ is 
a small-amplitude trigonometric profile. 
This initial small perturbation of the conductive state grows in time
and saturates as the stable traveling wave state. 
The frequency is initially $\omegacond$, but 
decreases as the amplitude grows, demonstrating 
the over-prediction of $\omegacond$ relative to $\omegatwobs$. 
The phase portraits resemble analogous ones by Barkley~\cite{Barkley} 
and by Noack~\cite{Noack} for the cylinder wake. 
Figure \ref{fig:TWfields}(e) plots $\Theta$
at two points on the midline separated by $\domain/4$.
The trajectory spirals out to the final saturated limit cycle.
A second phase portrait in figure \ref{fig:TWfields}(f) has 
as its vertical axis 
\begin{align}
P_M\equiv\int_0^{\domain}\:dx\: [\Temp(x,z=3/4)-\Temp(x,z=1/4)]
\label{eq:meanproxy}\end{align}
which approximates a projection onto the mean field.
This phase portrait shows clearly that the traveling wave 
does not orbit around the conductive state, which is at the bottom 
of the figure, but instead around the mean field $\bUb$,
whose projection onto this phase plane is located at the top. 

\section{Standing waves}
\label{sec:SW}

We now consider the standing waves that arises simultaneously with the
traveling waves at the Hopf bifurcation and seen in figure \ref{fig:bifdiag}.
Although they are unstable, standing waves can be calculated 
by direct simulation by imposing reflection symmetry in $x$.
Figure \ref{fig:SWfields}(a) shows instantaneous fields 
over one oscillation of a standing wave.
The mean field $\bUb_\SW$ is shown in figure \ref{fig:SWfields}(b).
As was the case for the traveling wave, the mean field bears 
little resemblance to the instantaneous standing waves.
However, contrary to those of the traveling wave, 
all of the SW mean fields vary in $x$.
The primary horizontal wavenumber seen in figure \ref{fig:SWfields}(a) is $k$, 
while those seen in figure \ref{fig:SWfields}(b) are 0 
for $\overline{\Temp}$ and $\overline{\Conc}$ 
and 0 and $2k$ for $\overline{\Psi}$. 
Although the mean streamfunction is not as small as in the TW case, 
we have $||\overline{\Psi}||/||\Psi||\approx 4\times 10^{-3}$ compared to 
$||\overline{\Temp}||/||\Temp||\approx||\overline{\Conc}||/||\Conc||=0.25$. 

The frequency $\omegaswobs$ of the standing waves and the leading eigenvalues 
of $\mL_\SW$ are shown in figure \ref{fig:SWfields}(c).
These behave quite differently from their traveling wave counterparts
in figure \ref{fig:TWfields}(c). Here, $\omegaswobs$ increases with $r$ near 
onset and remains quite close to the frequency $\omegacond$ 
of the conductive state 
up to $r=2.4$, while, in contrast, the frequency $\omegaswmean$ decreases
immediately after onset. 
However, for $r\gtrsim 2.4$, $\omegaswobs$ begins to diverge from $\omegacond$, 
veering down quite sharply starting at $r\sim 2.6$ and then less abruptly 
at $r\sim 2.8$, mirroring the behavior of $\omegaswmean$.
The distance between $\omegaswobs$ and $\omegaswmean$
remains finite and approximately constant
as the two move in tandem. In contrast $\omegacond$
continues to increase, thus diverging from $\omegaswobs$.
Thus, although $\omegacond$ is a much better predictor of $\omegaswobs$
for $2.1 \leq r\lesssim 2.4$, 
the downward trend of $\omegaswobs$ for $2.4 \lesssim r \leq 3$.
is tracked quite accurately by $\omegaswmean$.

In contrast to the traveling-wave case,
the growth rate $\sigmaswmean$ 
is positive and grows linearly along with that of the conductive state;
$\bU_\bSW$ is not a marginally stable state.
For $r\gtrsim 2.4$, $\sigmacond$ continues to increase 
linearly, while $\omegaswmean$ increases more steeply
for $2.6\lesssim r \lesssim 2.8$ and then tapers off for 
$2.8 \lesssim r \lesssim 3$.
Thus, the approach of $\omegaswmean$ towards $\omegaswobs$
for $r\gtrsim 2.4$ is not matched by an 
approach of $\sigmaswmean$ towards zero. 
In summary, the standing waves in thermosolutal convection do not have
the RZIF property, as manifested in the figure by the fact that the
long-dashed red and solid black lines are far from one another. The
real part of the mean-field eigenvalue is far from zero and the
frequency does not match the observed nonlinear frequency.

Figure \ref{fig:SWfields}(d)-(f) illustrates the dynamics of the approach 
to the standing wave, via a timeseries and two phase portraits. 
Figure \ref{fig:SWfields}(d), a timeseries of the mean-field 
projection proxy \eqref{eq:meanproxy},
shows a brief oscillation about the conductive state 
followed by saturated oscillations about the mean field.
The phase portrait in Fig.~\ref{fig:SWfields}(e) shows the 
instantaneous temperature and concentration at the domain midpoint; 
in this projection, the standing wave traces out a trapezoid, 
oriented along the positive diagonal, corresponding to the 
$T/8$ temporal phase shift between the two fields.
The phase portrait in Fig.~\ref{fig:SWfields}(f) demonstrates 
the difference between the standing and traveling wave dynamics:
the projection onto the mean field is far from constant. 
An important second harmonic component is visible in 
Fig.~\ref{fig:SWfields}(f), since 
a maximum in \eqref{eq:meanproxy} corresponds alternately 
to a minimum and a maximum in $C(\lambda/2,1/2)$.

\section{Temporal spectra}

We now present a simple but general analysis giving conditions 
under which the RZIF property is guaranteed.
Consider any evolution equation
\begin{align}
\pd_t \bU &=\mL \bU + \mN(\bU,\bU)
\label{eq:sys}\end{align}
where $\mL$ is linear and $\mN(\cdot,\cdot)$ is a quadratic nonlinearity.
Let
\begin{align}
\bU = \bUb + \sum_{n\neq 0} \bu_n e^{in\omega t}
\label{eq:periodic}\end{align}
(with $\bu_{-n}=\bu_n^\ast$)
be the temporal Fourier decomposition of
a periodic solution to \eqref{eq:sys} with mean $\bUb$ and 
frequency $\omega$. 
Substituting \eqref{eq:periodic} into \eqref{eq:sys} leads to 
\begin{align}
&\sum_{n\neq 0} in\omega\bu_n e^{in\omega t} = \mL\bUb 
+ \mN(\bUb,\bUb) + \sum_{m\neq 0} \mN(\bu_m,\bu_{-m})\nonumber\\
& + \sum_{n\neq 0}\mL\bu_n e^{in\omega t} + \sum_{n\neq 0}\left(\mN(\bUb,\bu_n) + \mN(\bu_n,\bUb)\right)e^{in\omega t} 
\nonumber\\& 
+ \sum_{n\neq 0}\sum_{m\neq 0,n}\mN(\bu_m,\bu_{n-m})e^{in\omega t} \label{eq:whole} 
\end{align}
Separating \eqref{eq:whole} by frequency leads to an equation for the zero 
frequency terms:
\begin{align}
0 &= \mL\bUb + \mN(\bUb,\bUb) + \sum_{m\neq 0}\mN(\bu_m,\bu_{-m})
\label{eq:zero}\end{align}
which is the same as Eq.~\eqref{eq:RANS} satisfied by the mean field, and
equations for the nonzero frequency terms ($n\geq 1$):
\begin{align}
in\omega \bu_n &= \mL \bu_n + \mN(\bUb,\bu_n) + \mN(\bu_n,\bUb) \nonumber \\
& + \sum_{m\neq 0,n}\mN(\bu_m,\bu_{n-m})
\end{align}
The first three terms on the right-hand-side of this expression are just 
those defining the linearization about $\bUb$. Hence, for 
$n\geq 1$ we have
\begin{align}
in\omega \bu_n &= \mL_\bUb \bu_n + \mN_n \label{eq:final}
\end{align}
where $\mN_n \equiv \sum_{m\neq 0,n}\mN(\bu_m,\bu_{n-m})$. 
For example:
\begin{subequations}
\begin{align}
\mN_1 &= \mN(\bu_2,\bu_{-1}) + \mN(\bu_{-1},\bu_2) \nonumber\\
&\qquad+ \mN(\bu_3,\bu_{-2}) + \mN(\bu_{-2},\bu_3) + \ldots \label{eq:n1}\\
\mN_2 &= \mN(\bu_1,\bu_1) + \mN(\bu_3,\bu_{-1}) + \mN(\bu_{-1},\bu_3) \nonumber\\
&\qquad + \mN(\bu_4,\bu_{-2}) + \mN(\bu_{-2},\bu_4) + \ldots \label{eq:n2}\\
\mN_3 &= \mN(\bu_1,\bu_2) + \mN(\bu_2,\bu_1) \nonumber\\
&\qquad + \mN(\bu_4,\bu_{-1}) + \mN(\bu_{-1},\bu_4) + \ldots \label{eq:n3}
\end{align}\label{eq:n}\end{subequations}
If $\bu_2 = \bu_3 = \ldots = 0$, i.e.~if the periodic cycle is 
exactly monochromatic, then $\mN_1=0$ and 
\begin{align}
i\omega \bu_1 = \mL_\bUb\bu_1 
\label{eq:good}\end{align}
i.e. $\mL_\bUb$ has an eigenvalue whose real part is zero and whose 
imaginary part is the frequency of the periodic solution. 
Hence the RZIF property necessarily follows for monochromatic oscillations 
in a system with quadratic nonlinearity.

More generally if, as is often the case, 
\begin{align}
||u_n||\sim \epsilon^{|n|}
\label{eq:order}\end{align}
then
\begin{align}
\underbrace{i \omega \bu_1}_{\epsilon} - \underbrace{\mL_\bUb\bu_1}_{\epsilon} = \underbrace{\mN_1}_{\epsilon^3}
\label{eq:approx}\end{align}
so that \eqref{eq:good} holds approximately.

The validity of \eqref{eq:good} or \eqref{eq:approx} 
does not require that $\mN_n$ 
be negligible in \eqref{eq:final} for higher harmonics $n>1$ 
and indeed it is not.  Assuming the scaling \eqref{eq:order}, 
equations \eqref{eq:n} show that for $n\geq 2$, 
$\mN_n$ is of the same order as the other terms in \eqref{eq:final}, e.g.
\begin{align}
\underbrace{i 2 \omega \bu_2 - \mL_\bUb\bu_2}_{\epsilon^2} = \underbrace{\mN_2}_{\epsilon^2},\qquad
\underbrace{i 3 \omega \bu_3 - \mL_\bUb\bu_3}_{\epsilon^3} = \underbrace{\mN_3}_{\epsilon^3}
\end{align}

\begin{figure}
\includegraphics[width=\columnwidth,clip]{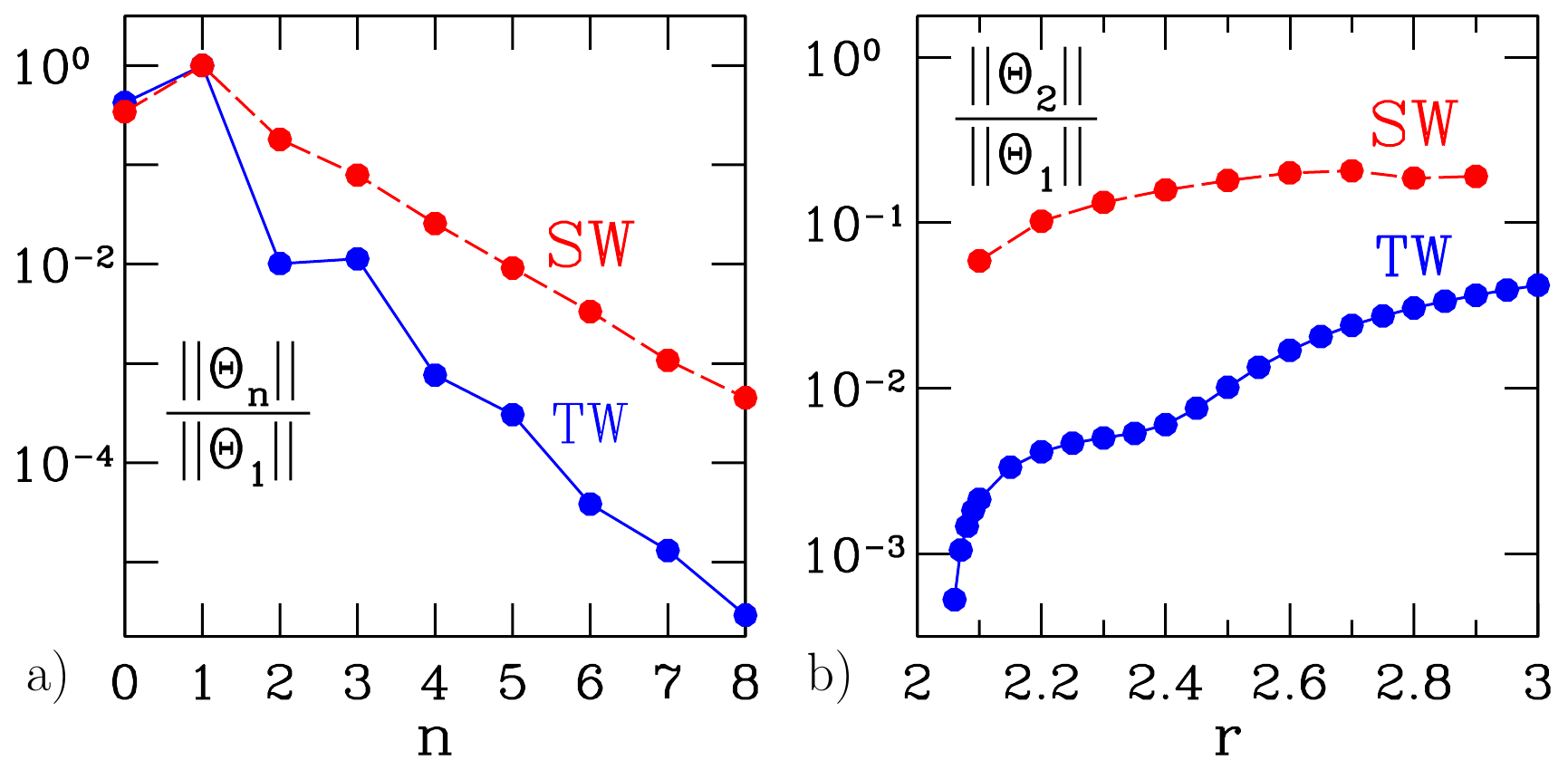}
\caption{(Color online) Temporal spectra of temperature field of
  traveling and standing waves.  (a) Full temporal spectrum for
  $r=2.5$, i.e. $||\Theta_n||$ for multiples $n$ of the observed
  frequency. (b) Ratio $||\Theta_2||/||\Theta_1||$ as a function of
  $r$.  The spectrum of the traveling waves (blue, solid) is far more
  peaked at $n=1$ than that of the standing waves (red, dashed).  }
\label{fig:spectra}
\end{figure}



Although \eqref{eq:good} is linear in $\bu_1$, the mean flow equation
\eqref{eq:zero} is quadratic in $\bu_1$, insuring the saturation of
its amplitude, as shown more explicitly in system \eqref{eq:Gallaire}
below.  Quadratic interaction of $\bu_1$ with itself takes two forms:
$\mN(\bu_1,\bu_{-1})+\mN(\bu_{-1},\bu_1)$ leads the mean flow $\bUb$ to
differ from the base flow, while
$\mN(\bu_1,\bu_1)$ generates $\bu_2$.  RZIF is favored by small
$\bu_2$, and hence by small $\mN(\bu_1,\bu_1)$.  On the other hand,
since RZIF requires a mean field which differs from the base
field, $\mN(\bu_1,\bu_{-1})+\mN(\bu_{-1},\bu_1)$ should be
non-negligible.  Nonlinear saturation of amplitudes of instabilities
in a more usual non-RZIF context is addressed in texts
such as Manneville \cite{Manneville} and Iooss and Joseph \cite{Iooss_Joseph}.

The reasoning above does not provide an {\it a priori} reason for 
which an eigenvalue of the mean field would predict the frequency, 
since no method has been given for determining whether the 
oscillatory state is almost monochromatic. 
It does, however, connect these two properties.
The frequency of a monochromatic or almost-monochromatic 
oscillation in a system with quadratic nonlinearity should be well 
predicted by the leading eigenvalue of $\mL_\bUb$.


Returning to the thermosolutal system, 
figure \ref{fig:spectra} contrasts the temporal spectra of the 
temperature field of the traveling and standing waves. 
Figure \ref{fig:spectra}(a).
shows that the spectrum of the traveling wave at $r=2.5$ 
is concentrated at $n=1$, i.e.~at the observed frequency. 
The spectrum of the standing waves is far wider, with substantial 
amplitude for $n\geq 2$.
The ratio $||\Temp_2||/||\Temp_1||$ is $10^{-2}$ for TW 
and 20 times higher than this for SW.

Figure \ref{fig:spectra}(b) shows that this trend holds 
over our range $2\leq r \leq 3$ of observation, 
by plotting $||\Temp_2||/||\Temp_1||$ as a function of $r$.
Interestingly, for TW this ratio shows an upturn 
at $r\approx 2.5$, which is where 
figure \ref{fig:TWfields}(c) shows that 
$\sigma^{\rm TW}_{\rm mean} + i\omega^{\rm TW}_{\rm mean}$ 
begins to deviate from $0+ i \omega^{\rm TW}_{\rm exact}$. 
For SW, there seems to be little or no correlation between the 
$r$-dependence of the temporal spectrum and that of 
$\sigma^{\rm SW}_{\rm mean} + i\omega^{\rm SW}_{\rm mean}$, 
which is to be expected if the relationship between 
the two requires a peaked spectrum.

The RZIF property is corroborated by order-of-magnitude comparison of
the terms in equations \eqref{eq:approx}.  Examining the equations
corresponding to the temperature field for TW at $r=2.5$, the ratio of
the maximum of the right-hand-side (which we wish to neglect) to that
of the left-hand-side is 0.025.  Comparing the quadratic terms
which appear in the temperature equation, we find that the ratio between 
$\mJ[\Psi_1,\Temp_1]$ and $\mJ[\Psi_1,\Temp_{-1}]+\mJ[\Psi_{-1},\Temp_1]$
is 0.1.

The spectra of the other fields comprising ${\bf U}$ also 
follow this tendency, but not to the dramatic extent of $\Theta$.
The concentration field at $r=2.5$ has 
$||\Conc_2||/||\Conc_1||\approx 0.1$ for TW
(ten times larger than for $\Theta$) and 0.2 for SW
(the same as for $\Theta$).
Similarly, for the TW concentration field,
the ratio of the maximum of the right-hand-side of \eqref{eq:approx}
to that of the left-hand-side is 0.2 (ten times this ratio for $\Theta$).
The ratio between the quadratic terms which appear in the concentration
equation, $\mJ[\Psi_1,\Conc_1]$ and
$\mJ[\Psi_1,\Conc_{-1}]+\mJ[\Psi_{-1},\Conc_1]$, is 0.4.

\section{Relation to previous work on cylinder wake}

In light of these results, 
we review some of the previous work concerning RZIF.
Linear stability analysis of the mean field has been 
carried out only for open flows and almost exclusively for the 
cylinder wake. 
Hammond and Redekopp \cite{Redekopp}, Pier \cite{Pier}, 
Barkley \cite{Barkley} and Mittal \cite{Mittal} 
have shown that the frequency of the cylinder wake is predicted 
with remarkable accuracy by that of the mean field 
even quite far above onset, 
at least until $Re=180\approx 4 Re_{\rm H}$. 
In keeping with our conjecture that RZIF coincides with a monochromatic 
oscillation, Dusek et al.~\cite{Dusek} have found experimentally that 
the temporal spectrum is highly peaked even at high Reynolds numbers.
Knobloch et al.~\cite{Knobloch_89} find that the 
traveling waves in a minimal model of thermosolutal convection 
are almost monochromatic, while the standing waves at the same parameter 
values are not. 

Sipp and Lebedev~\cite{Sipp} carried out a numerical weakly nonlinear 
analysis of the cylinder wake about the Hopf bifurcation point, 
and were able to reproduce the slope of the frequency as well as 
the zero growth rate. 
More specifically, they approximated the flow $\bU$ and 
its frequency $\omegaobs$, its mean flow $\bUb$ and 
the eigenvalues of this mean flow 
$\sigmamean \pm i \omegamean$ near onset $Re_{\rm H}$ and 
found agreement between the slopes of the mean field frequency and 
the nonlinear frequency $\omegamean\p(Re_H)\approx \omegaobs\p(Re_H)$ 
as well as marginal stability $\sigmamean\approx 0$.
They did not capture its further evolution with $Re$, for 
example its curvature at $Re_{\rm H}$,
which would have required extending the analysis to higher order. 

More fundamentally, Sipp and Lebedev~\cite{Sipp} 
determined which of the contributions to the lowest-order nonlinear term 
were required to be small in order to achieve this agreement.
These terms arise from the second temporal harmonic, 
just as we have found.
They also presented an important counter-example.
Performing the same weakly nonlinear analysis 
on the flow in an open driven cavity, they found both that the second 
harmonic contributions to the nonlinear term were not small 
and also that the weakly nonlinear analysis did not 
reproduce $\omegaobs\p(Re_H)$. 
This counter-example shows that RZIF does not hold for all flows 
and also corroborates the role of the second temporal harmonic. 

Our attempt to carry out a weakly nonlinear analysis 
analogous to that of Sipp and Lebedev~\cite{Sipp} was 
hampered by the degeneracy of the Hopf bifurcation to traveling waves 
in thermosolutal convection with free-slip boundary 
conditions \cite{Knobloch_85}; see figure \ref{fig:bifdiag}.
The cubic term in the normal form is zero, requiring the calculation of 
a quintic term or an appropriate model \cite{Knobloch_89}.
The bifurcation to standing waves is free from this pathology, 
but, as shown in section \ref{sec:SW}, the mean fields of the 
standing waves do not have the desired property near onset. 

Manti\v{c}-Lugo et al. \cite{Gallaire} proposed the following system 
\begin{subequations}\begin{align}
0 &= \mL\bUb + \mN(\bUb,\bUb) + A^2\mN(\bu_1,\bu_{-1}) \label{eq:Gal1}\\
(\sigma + i \omega) \bu_1 &= \mL_\bUb \bu_1
\qquad\qquad ||u_1||=1\label{eq:Gal2}\\
\sigma &= 0 \label{eq:Gal3}
\end{align}\label{eq:Gallaire}\end{subequations}
as a means of calculating $\bUb$ and $\omega$ for the cylinder wake 
without recourse to time integration.
Equation \eqref{eq:Gal1} is a truncated version of 
the exact equation \eqref{eq:zero} satisfied by the mean field;
consequently, its validity requires a stongly peaked spectrum,
as emphasized by Manti\v{c}-Lugo et al. \cite{Gallaire}.
Starting from an estimate of the mean field by the base flow, 
their equivalent of the conductive state, 
they solved \eqref{eq:Gal2} for the eigenvalue $(\sigma + i \omega)$
and the eigenvector $\bu_1$. 
Substituting $\bu_1$ in \eqref{eq:Gal1}, they computed a new 
mean field $\bUb$. The amplitude $A$ multiplying the eigenvector
was adjusted until convergence to marginal stability $\sigma=0$. 
Our attempt to carry out this iterative procedure for the thermosolutal 
system did not converge. 

\section{Conclusion}

\begin{figure}
\includegraphics[width=\columnwidth]{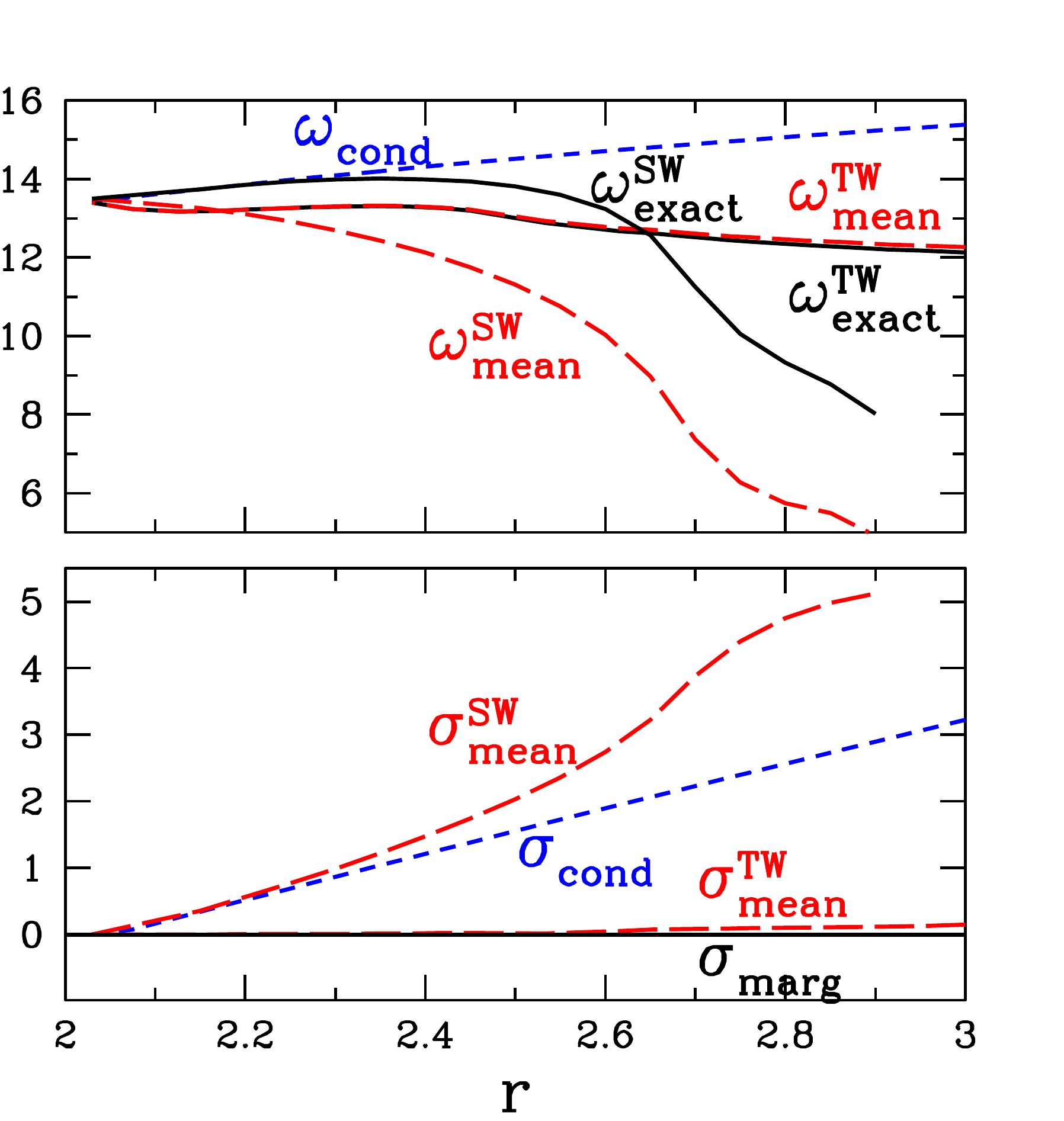}
\caption{(Color online) Frequencies (above) and growth rates (below) for TW and SW.
The mean eigenvalues (long-dashed red) track the exact frequency 
and marginal growth rate (solid black) very closely for TW 
and not at all for SW.}
\label{fig:botheigs}
\end{figure}

A number of fluid-dynamical researchers have attempted to 
relate the nonlinear frequency of a periodic state, 
primarily the cylinder wake, with 
the imaginary part of the eigenvalue of the evolution operator 
linearized about the temporal mean. 
Following this line of investigation, we have studied 
the traveling waves and standing waves 
resulting from the Hopf bifurcation in thermosolutal convection. 
The traveling waves have turned out to be a textbook case of RZIF, i.e. the 
real part of the mean-flow eigenvalue is almost exactly zero 
and the imaginary part is almost exactly the frequency.
In contrast the standing waves do not have this property:
the mean-flow eigenvalue performs even worse than the 
base-flow eigenvalue at predicting the frequency. 
These results are displayed in our summary diagram, 
figure \ref{fig:botheigs}. 

Guided by these results, we have 
put forth a general theoretical explanation for the RZIF 
property in terms of the temporal spectrum. 
If the periodic oscillation is monochromatic, then RZIF is satisfied exactly. 
If it is not exactly monochromatic but the higher 
temporal harmonics are small, then RZIF 
should be satisfied approximately. 
This corresponds to the traveling/standing wave dichotomy:
the spectrum of the traveling waves is sharply peaked, 
while that of the standing waves is not. 
The question which remains is that of predicting the 
width of the spectrum of periodic states. 

\acknowledgments

We acknowledge funding from the Agence Nationale de Recherche (ANR)
for the TRANSFLOW project and Pembroke College of Cambridge
University, as well as discussions with Alastair Rucklidge on this
topic at an earlier stage of this work.

\appendix

\section{Numerical Methods}

Because of the horizontally periodic and vertical no-slip boundary conditions 
\eqref{eq:bcs}, the fields can be represented as 
\begin{align}
\bU(x,z,t) = \sum_n \hat{\bU}_{mn}(t) e^{imkx}\sin(n\pi z)
\end{align}
for which spatial derivatives are easily taken.
A two-dimensional Fourier transform leads to values on an equally 
spaced grid, where the multiplications in the Poisson bracket $\mJ$ 
of \eqref{eq:Poisson} are performed. 

In order to solve 
\begin{align}
\pd_t\bU &= \mL\bU(t) +\mN(\bU,\bU) \label{eq:compact2}
\end{align}
we use first-order explicit-implicit time discretization, i.e.
backwards Euler for the diffusive terms and 
forwards Euler for all other terms. 
\begin{align}
\bU(t+\dt) &= (I-\dt \mL)^{-1} (I + \dt \mN)\bU(t) \label{eq:FEBE}\\
&\equiv B\bU(t)\nonumber
\end{align}

In order to explain the method by which traveling waves are calculated, 
we begin by discussing the calculation of steady states. 
We adapt the timestepping scheme 
\eqref{eq:FEBE} to carry out Newton's method as follows \cite{Mamun}:
\begin{align}
\bU(t+\dt) - \bU(t) = &\left [\left(I-\dt\mL \right)^{-1} \left(I +\dt\mN \right) - I \right]\bU(t)\nonumber\\
= \left(I-\dt\mL \right)^{-1} &\left[I + \dt\mN - \left(I-\dt\mL\right)\right]\bU(t)\nonumber\\
= \left(I-\dt\mL \right)^{-1} &\dt \left(\mN+\mL\right)\bU(t)
\label{eq:equiv}\end{align}

Thus, the roots of $\mN+\mL$ are the same as those of 
$B-I$, where $B$ is the timestepping operator \eqref{eq:FEBE}.
The calculation \eqref{eq:equiv} holds for any value of $\dt$ 
regardless of size,
in contrast to \eqref{eq:FEBE}, whose validity as 
a timestepping scheme relies on a Taylor series 
approximation in $\dt$ and so requires $\dt$ small.
One Newton step is then formulated as 
\begin{align}
(B_\bU-I) \bu &= (B-I) (\bU) \label{eq:newton}\\
\bU &\leftarrow \bU - \bu \nonumber
\end{align}
where
\begin{align}
B_\bU \bu &\equiv (I-\dt \mL)^{-1} (\bu + \dt(\mN(\bU,\bu)+\mN(\bu,\bU))
\end{align}
is the linearization of $B$ about the current solution estimate $\bU$
and $\bu$ is the decrement to be calculated and applied to $\bU$.
For large $\dt$, the linear operator $B_\bU-I$ is well conditioned, 
unlike the usual Jacobian operator $\mN_\bU + \mL$.
BICGSTAB \cite{Vandervorst} is used to solve the linear 
equation \eqref{eq:newton} for $\bu$, thus avoiding
the storage and even the construction of the Jacobian operator. 

In the horizontally periodic domain, solutions are not unique but 
defined only up to a spatial phase; equivalently, the Jacobian is 
singular, with the marginal $x$-translation of $\bU$ as a null vector. 
(Although the image of the Jacobian $B_{\bU}-I$ is not of full rank, 
a vector produced by action of $B-I$ is always in its image.)
A singular linear system can, however, be solved by iterative 
conjugate gradient methods without imposing any additional phase condition,
since the solution is constructed from a set of vectors 
created by repeated action of the linear operator on an initial vector, 
which is unaffected by the singularity of the Jacobian.
The solution $\bu$ returned by BICGSTAB will be one of the 
possible solutions of \eqref{eq:newton}, whose phase is 
determined by that of the right-hand-side $\bU$.

Traveling wave solutions are of the form
\begin{align}
\bU(x,z,t) &= \bU(x-Vt,z) \label{eq:TW}
\end{align}
where $V$ is the unknown velocity.
Substituting \eqref{eq:TW} into the governing equation \eqref{eq:compact2}
leads to 
\begin{align}
0 &= \mL \bU + \mN(\bU,\bU) + V \pd_x\bU 
\label{eq:steady}\end{align}
so that $(\bU, V)$ together describe a steady state.
We redefine $B$ and its linearization as:
\begin{align}
&B (\bU,V) = (I-\dt \mL)^{-1} [\bU + \dt(\mN(\bU,\bU) + V\pd_x\bU)] \\
&B_{\bU,V} (\bu,v) = (I-\dt \mL)^{-1} \nonumber\\
&\qquad [\bu +\dt( \mN(\bU,\bu)+\mN(\bu,\bU)+ V\pd_x\bu - v\pd_x \bU)]
\end{align}
For any periodic orbit the solution is defined only up to a temporal phase.
Unlike in the case of a spatial phase,
here it is necessary to impose an additional equation 
since $V$ is an additional unknown. 
The easiest option is to fix the zero-crossing
of one of the elements of $\bU$ 
by specifying that it remain unchanged by the Newton step. 
(Since the fields we compute are deviations from the linear profile, 
all of them cross zero at any height $z$; any such zero-crossing 
can be set as a condition, i.e. $u_p=0$.)
Since $u_p$ is to be set to zero, 
a computational simplification can be realized 
by storing $v$ in this element. The subroutine corresponding to 
the action of $B_{\bU,V}$ begins by extracting $v$ from 
$u_p$, replacing $u_p$ by its imposed value of zero:
\begin{align}\begin{array}{cl}
v &\leftarrow u_p\\
u_p &\leftarrow 0
\end{array}\end{align}
and then acting with $B_{\bU,V}$. 
This procedure is repeated when a solution is returned by BICGSTAB
before decrementing:
\begin{align}\begin{array}{cl}
\bU &\leftarrow \bU - \bu\\
V &\leftarrow V - v
\end{array}\end{align}
Standing waves, which cannot be calculated in this way, 
are computed by integrating in time while imposing reflection symmetry 
in $x$.

We calculate leading eigenvalues by constructing and then 
diagonalizing the Jacobian, restricting computation to the 
horizontal wavenumber $k$. 


\end{document}